\documentclass[notitlepage]{article}%
\usepackage{amsmath}
\usepackage{amsfonts}
\usepackage{amssymb}
\usepackage{graphicx}
\usepackage{amsthm}
\usepackage{lineno}
\usepackage[nottoc]{tocbibind}
\usepackage{endnotes}

\begin{document}
\bibliographystyle{plain} 
\pagestyle{myheadings}
\markright{Lexical growth, entropy and the benefits of networking }

\title{Lexical growth, entropy and the benefits of networking } 
\author{Robert Shour}\
\date{}
\begin{center}
\ {\Large\textbf\ Lexical growth, entropy and the benefits of networking } \\
\normalsize
\hfill \\
\ Robert Shour
\hfill \\
\hfill \\
\footnotesize{Toronto, Canada}
\end{center}
\hfill
\normalsize
\begin{abstract}
\noindent
If each node of an idealized network has an equal capacity to efficiently exchange benefits, then the network's capacity to use energy is scaled by the average amount of energy required to connect any two of its nodes. The scaling factor equals \textit{e}, and the network's entropy is $\ln(n)$. Networking emerges in consequence of nodes minimizing the ratio of their energy use to the benefits obtained for such use, and their connectability. Networking leads to nested hierarchical clustering, which multiplies a network's capacity to use its energy to benefit its nodes. Network entropy multiplies a node's capacity. For a real network in which the nodes have the capacity to exchange benefits, network entropy may be estimated as $C \log_L(n)$, where the base of the log is the path length $L$, and $C$ is the clustering coefficient. Since $n$, $L$ and $C$ can be calculated for real networks, network entropy for real networks can be calculated and can reveal aspects of emergence and also of economic, biological, conceptual and other networks, such as the relationship between rates of lexical growth and divergence, and the economic benefit of adding customers to a commercial communications network. \textit{Entropy dating} can help estimate the age of network processes, such as the growth of hierarchical society and of language.
\end{abstract}
\



\setlength{\unitlength}{4in}
\begin{picture}(.1, .1)
 \put(.0625, 0){\line(1, 0){1}}
 \end{picture}

\hfill\\


What are the benefits of networking for a person and for a lexicon? More precisely, and more generally, how much is a process's nodal rate multiplied by networking?

Language growth raises the above questions. To grow a language, a society must perform three problem-solving processes: 
\begin{enumerate}
	\item Devise sounds (phonemes) and choose which ones will be used and clustered for coding. 
	\item Identify, conceptualize and choose which perceptions, and which abstractions arising from them, should be coded. 
	\item Decide, using the chosen coding sounds, how to code and cluster chosen perceptions and abstractions. 
\end{enumerate}

Language memorializes the products of these three problem-solving processes. As a society ages, the lexicon grows, the emergent product of problem solving. Consistent with this perspective, the English lexicon grew an average of about 3.4\% per decade from 1657 to 1989\endnote{ The Early Modern English Dictionaries Database (EMEDD) at the University of Toronto, $www.chass.utoronto.ca/english/emed/ \#dic$ at October 15, 1999 had about 200,000 word-entries at 1657. The Oxford English Dictionary (OED) at 1989 had about 616,000 word-entries. Word-counts vary among dictionaries. I assume that lexical criteria are similar for these sources. Because of the recency of historical dictionary projects, lexical growth as a metric of language appears not to have been previously considered.}, and average IQs, an approximate measure of society's problem solving skill, grew at about the same rate in the U.S.A from 1947 to 2002\endnote{ J. R. Flynn, Journal of Educational Measurement, \textbf{21}(3), (1984), 283; Psychological Bulletin, \textbf{101}(2) 171 (1987); \emph{What is Intelligence?} Cambridge University Press, 2007.}. Since lexicon formation and growth is an emergent process in a society, the similarity in the rates of growth is consistent with increasing average IQs also being an emergent process. This implies that the rate at which society emergently improves its capacity to solve problems, language being a particular accomplishment of society's problem solving capacity, may possibly be measured indirectly by measuring lexical growth. 

As the number of people in society grows, society's capacity to invent words increases: the lexical growth rate increases. If we could quantify how much society on average multiplies an individual's capacity for lexical growth, we might then be able to calculate an average basic lexical growth rate, and use that rate as a clock to estimate when language began.

What is the benefit of networking?

Consider the position of a child dependent on society for information. A child receives information directly from $L$ sources, including parents and close friends. Parents in turn receive information from the child's four grandparents, eight grandparents and so on. Each of the $L$ direct sources receives information from (as a simplification) the same average number of $L$ sources. This provisionally suggests that the multiplicative benefit of networking for the child equals a log function, $\log_L(n)$, where $n$ is the number of people in the society. Only decreasing $L$ is consistent with both increasing the value of the log and increasing the network benefit: to increase the network's information benefit, the base of the log must decrease. Hence, $L$ must be proportional to an average time or distance to the information sources because reducing the average time or distance to connect to information sources would increase the rate of information transmitted to a recipient. If the network benefit $H$ is $\log(n) = \eta$ with $L$ as the base of the log (equivalently, $L^\eta = n$), one may infer that the network is hierarchical (from $L^\eta$), that, likewise, $L$ is the scaling factor, that the nodes must all have equal (or the same average) attributes since the formula for network benefit does not distinguish between nodes, and that the hierarchy must be flat, since $L^\eta = n$ requires that lower levels in the hierarchy contain the same $n$ nodes. The average distance (or time) between nodes is non-commensurable with the number of nodes. We seek $L$ commensurable with a parameter of the entire network. If energy is proportional to the distance a signal has to traverse, then the average energy for the reception of information from another node in the network scales the (commensurable) energy of the network. Adjacency plays a critical role since a node cannot connect to non-adjacent nodes without first connecting with an adjacent node. 

The foregoing observations guide the modeling of an ideal network $N$ with the following attributes.

\begin{enumerate}
	\item  $N$ exists. Its nodes require energy, are connectable and can transmit and receive benefits. $N$ has n different but otherwise indistinguishable nodes, where n is greater than one and finite. Each node has the capacity to transmit and receive benefits. Each node in $N$ has some adjacent nodes, and all other nodes are non-adjacent. A pair of nodes are adjacent if they are connectable in one step, and are non-adjacent if they are only connectable in multiples of one step. Each one step connection only needs to be created once. Each creation of a one step connection has the same finite energy cost. Each node's energy is continuously supplied, at a finite rate, solely by its environment. At every point in time, energy units are defined so that one unit of energy per unit of time transmits one unit of benefit one step. For a succession of network equilibrium states, the energy units and time units are adjusted if necessary to maintain their one-to-one proportion to each other.
	\item Every node in $N$ can respond to its environment and will minimize its use of energy for the acquisition of each unit of benefit received from the environment or from another node in $N$, and will maximize the benefits it receives for each unit of energy it expends.  This attribute may be called nodal self-interest.
\end{enumerate}

The next 11 propositions follow from the preceding attributes: 

\begin{enumerate}
	\item Nodes have the capacity to transmit multiple benefits: because energy is continuously supplied to them.
	\item Nodes connect. When the benefit received is greater than the energy cost, a node connects to another node, because of nodal connectability and self-interest. Even if the energy cost of a connection exceeds the value of the benefit, the benefit of receiving multiple transmissions will at some point exceed the cost of connecting, because connecting has a one-time finite cost.
	\item Adjacent nodes connect in one step because nodes maximize the benefit per unit of energy, and the benefit of connecting to adjacent nodes per unit of energy is higher than the benefit of connecting to non-adjacent nodes per unit of energy. There is an energy advantage to adjacency and nearer proximity. 
	\item Nodes connect bi-directionally when possible: because one bi-directional connection costs less energy to create than two single direction connections, and nodes can both transmit and receive; all single step connections have the same energy cost.
	\item An average number of steps $L$ between pairs of nodes in $N$ exists: because the number of nodes and, therefore, the number of steps between nodes, are each finite. Since non-adjacent nodes exist for every node, $L$ is a positive number greater than $1$. $L$ plays a critical role in the emergence of a network, as discussed later in these propositions and later in this paper.
	\item The average number of energy units to transmit one unit of benefit from one node to another is $L$: because one unit of energy transmits one unit of benefit one step in one unit of time, and $L$ is the average distance in steps between pairs of nodes. Therefore, at every point in time, $L$, in addition to being the average number of steps between nodes, is also the average number of energy units per unit of time required by a node to transmit a benefit to another node. $L$ is proportional to the average energy required to connect two nodes.
	\item $N$ uses $n$ units of energy per unit of time, at each point of time: because of the way energy units are defined at each point in time.
	\item $N$'s energy use is scaled by $L$ at each point of time. Suppose, to simplify calculation, an external energy source continuously transmits energy to each node in $N$ at a constant rate, so the ratio of the rate of energy units transmitted to the benefit per step is constant. Suppose the energy source, the $zero^{th}$ energy generation, is one step away\endnote{ So no node has a preferred role as a transmitter. All $n$ nodes are potential recipients.} from $N$ and transmits energy benefits to a single path of $L$ steps, thereby reaching $L$ first generation nodes. 

Suppose further that each recipient first generation node retransmits energy benefits along single paths, since every recipient node can also transmit. By the connectability of nodes and the existence of $L$, each node on the initial path of $L$ steps can transmit to $L$ nodes using $L$ energy units. $L^2$ second generation nodes receiving transmissions from $L$ first generation receive $L^2$ energy benefits. Each path of $L$ nodes can connect to $L$ times as many nodes, until the energy benefits reach all $L^\eta = n$ nodes in $N$, for some $\eta$. Now, instead, suppose the external source has the capacity to transmit energy benefits to all single paths of $L$ distinct nodes in $N$ (that is, $n/L$ such paths), and those first generation nodes have the capacity to transmit in turn as above, which is possible because all nodes are equally capable of receiving and transmitting. The number of all $\eta^{th}$ generation nodes cannot total more than $n$ distinct nodes. In view of $L$ being the average distance in steps between nodes, the $\eta^{th}$ generation nodes cannot be more than an average of $L$ steps away from $N$'s other nodes, even though $\eta$ can be larger than $L$. Nodes are also contained in generations preceding those in the $\eta^{th}$generation.

The only way this capacity for clustering can be accomplished is if the $n$ nodes in each generation are the same $n$ nodes as those in every $\eta^{th}$ generation of nodes, and if $k^{th}$ generation clusters are nested in $(k-1)^{st}$ generation clusters. It must be that every node that is a member of a cluster is also contained in a cluster closer to the energy source that is $L$ times larger, up to the $zero^{th}$ energy generation. The $L$ first generation clusters have size $L^{\eta-1}$, the $L^2$ second generation clusters have size $L^{\eta-2}$, and so on. The exponent in the exponential formula for the capacity to transmit benefits must be the same as the base of the log in the formula for the capacity to receive benefits, for $N$ to be scaled by $L$.

At every point in time, $N$'s receipt of $n$ energy units per unit of time is scaled by the average number of energy units used to traverse the average number of steps $L$ between nodes. Since connections are bi-directional, if a node can transmit benefits to $\eta$ cluster generations, it can also receive benefits from $\eta$ cluster generations. 

As a model of $N$'s capacity for scaling, consider 27 nodes in a single line forming a \textit{flattened} hierarchy scaled by 3. Differently scaled clusters are bracketed using differently shaped brackets: 

\begin{center}
{[(\ldots)(\ldots)(\ldots)][(\ldots)(\ldots)(\ldots)][(\ldots)(\ldots)(\ldots)]}.
\end{center}

In the model's 27 nodes, single nodes scale up to clusters of 3. Clusters of 3 scale up to clusters of 9. Clusters of 9 scale up to a cluster of 27. I infer that differently sized clusters have different emergent processes. The 27 nodes have three hierarchical cluster generations, though we can observe only one row of 27 nodes. 

The English language is structured in a similar way. For example, in its alphabetical representation, the English letters, \textit{s}, \textit{h}, and \textit{t} can combine to form \textit{sh} and \textit{th}, two-letter clusters that sound differently than their component letters. Differently sized letter clusters can join to form word roots, prefixes, suffixes, and larger clusters we call words. Words, together with spaces, can form noun, verb and object clusters, called phrases. Word clusters from different cluster generations can form sentences. But an alphabetically represented sentence just consists of a single string of individual letters and spaces. Similar observations apply to a sentence coded by a string of sounds. Society, using grammar emergently and hierarchically organizes language\endnote{ V. Fromkin and R. Rodman, An Introduction to Language (6th ed.) (Harcourt Brace, New York, 1998), p. 77, 111; A. Radford, M. Atkinson, D. Britain, H. Clahsan, A. Spencer, Linguistics - An Introduction (Cambridge University Press, Cambridge U.K., 1999), p. 88.}. If so, the 1950s hypothesis, still current, that there is a grammar module in the brain (a `language instinct'), is unnecessary. Grammar emerges through the adaptive network efficiency of a society using a lexicon. 

Reductionism applied to a conceptual problem involves the application of problem solving (energy) to a conceptual cluster that is a part of the larger problem.

The design of a place holding numeration system enables the combination of different cluster generations to describe numbers. The number $528$, for example, combines and contains five second generation clusters ($100$s), two first generation clusters ($10$s) and 8 $zero^{th}$ generation singletons. Each cluster size is a separate concept. 

Social networks, music\endnote{ D. J. Levitin, This is Your Brain on Music - The Science of a Human Obsession (Dutton, New York, 2006).}, laws\endnote{ H. Kelsen, Hans, Pure Theory of Law (The Lawbook Exchange, Clark, New Jersey, 2005).}, and athletic moves are also clustered hierarchically.
	\item $N$ is self-similar, since efficiency considerations applicable to adjacent nodes apply also to adjacent clusters. A cluster is part of a cluster $L$ times larger. For a node or cluster to efficiently maximize the benefits it receives, reception of benefits from a cluster smaller than N may be sufficient. For a recipient node or cluster to be efficient, it must not call upon more of the network's clustered energy resources (or energy capacity) than is minimally necessary for it to obtain the benefits it needs in particular circumstances, but it has the capacity to call on any of those clusters. For $N$ to be efficient, it must not use more of its energy resources to benefit a node or cluster than is necessary, but it has the capacity to transmit all of its resources to a node or cluster. Energy clustering enables efficient allocation of N's energy resources. For $N$ to be everywhere self-similar, there can be no local variations in the path length; the energy requirements per node must be \textit{everywhere} equal.
	\item The benefit of a network emerges: $\eta$ multiplies the capacity of a single recipient node. Since each of the $\eta$ generations in the hierarchy of clusters contains all n nodes in $N$, $N$ has the capacity to communicate $\eta$ times the capacity of one cluster generation to a node. (When $\eta = 0$, no energy units are transmitted because there is no energy to transmit. The capacity of a cluster of size $L$ is itself some multiple of the capacity of an individual node.) So if the measure of the capacity of a single generation of clusters containing all $n$ nodes to benefit a node is $A(L)$, the measure of the capacity of all $\eta$ cluster generations to benefit a node is $\eta A(L)$. But the capacity of all $\eta$ cluster generations is the capacity of the network, $A(n)$. Hence $A(n) = A(L^\eta) = \eta A(L)$, and $A$ must be a logarithmic function with base $L$. If $H_L$ represents the capacity of a network of $n$ nodes to benefit a node, $H_L(n) = \log_L(n) = \log_L(L^\eta) = \eta$. The capacity of $N$ to multiply the effect of its energy resources depends on clustering. A node can benefit only by receiving transmissions from a cluster, and a cluster can increase its capacity only if nodes and clusters transmit to it. Without the capacity for both reception and transmission\endnote{ Zipf discusses the efficiency conflict in language between speakers and hearers. G. K. Zipf, Human Behavior and the Principle of Least Effort (Hafner Publishing Company, New York, 1949, 1972 reprint).}, this could not occur, or at least would not necessarily occur in an efficient way, contrary to the assumption that nodes are benefit maximizers. A network can also temporally network with recorded earlier editions of itself, like a person proof-reading their own earlier work.

Nodal self-interest, combined with connectability, leads to the emergence of a network that benefits the network's nodes. 

Transmission by a social network of a social benefit to a recipient is an indirect transfer of the network's logarithmically compressed energy. A lexical network (language) logarithmically compresses the transfer of energy (the energy used to solve the problem of compressing perceptions into concepts) between and among the members of society who use the language. By receiving information expressed in words, a recipient can receive and share in benefits arising from the previous expenditure of energy by other members of society, past and present, which energy was used in one or more of the three problem solving processes involved in creating language. Because of networks, a member of society need not be adjacently connected to receive such benefits from a remotely connected other member of society, past or present.

	\item For an idealized network, $L = e$. For every node in the first generation, the number of nodes it has transmitted to increases from generation to generation at the rate $L$, until it has reached all the nodes of the $\eta^{th}$ generation. For continuous functions, if a function is its own derivative, so that $y' = y$, then $y = f(x) = e^x$, to which the behavior of the $L$-sized clusters is similar. The self-similarity of $N$ in all generations therefore implies that $L$, the base of the log, is the natural logarithm, namely $e$, about 2.71828. In that case, a network's benefit is $\ln(n)$. The optimal path length for a one-way broadcasting node is 1 (but such a node would require more energy than average to broadcast to the network). If nodes did not all have equal capacity for transmitting and receiving, then $N$ would not necessarily be self-similar in all cluster generations.
\end{enumerate}

To determine the network benefit for a node in an ideal network, the attributes above seem sufficient; microstates of the nodes and clusters are not of interest because the scaling factor is an average. In their seminal 1998 article\endnote{ D. J. Watts and S. H. Strogatz, Nature (London), \textbf{393}, 440 (1998).}, Watts and Strogatz use three parameters, $n$, $L$ and $C$, to characterize a kind of real network they call `a small world network'. The first parameter, $n$, is the number of nodes. $L$ is the path length, the smallest number, averaged over all pairs of nodes, of steps between nodes. $C$ is the clustering coefficient, the fraction of allowable edges, connecting to a vertex in a graph of the network, that actually exist, averaged over all nodes. The clustering coefficient can also be defined using the notion of adjacency. Suppose we calculate, for every node, the proportion of its adjacent nodes that are connected to it. The clustering coefficient, $C$, is the average of those proportions for $N$'s nodes. For a real network, the number of steps between nodes and the proportion of connected adjacent nodes are measured for all, or a representative sample, of the network's nodes, and the results are averaged to obtain $L$ and $C$. Long distance connections between clusters result in the `small world effect', sometimes described as `six degrees of separation'. 

For a real network, the clustering coefficient is between zero and one, which differs from an ideal network which implicitly assumes $C$ is 1. Thus for a real network, only a proportion C of the benefit of the network reaches a node, and for $n$, $L$ and $C$ at a given point in time,

\begin{equation}\label{Eq Net20.10}
	H_L(n) = C\log_L(n).
\end{equation}

In a real network, nodes might be unequal in capacities, energy requirements, and the number of steps between nodes. An average number of steps $L$ exists, however, because, whether for topological, physiological or other reasons, when the number of nodes is large, they cannot all bi-directionally connect to all other nodes in one step. In a real network, the fraction per step (energy/benefit) may differ from 1. For a network, $e$ is a benchmark. Suppose that for a real network, the per step fraction (energy/benefits) $< 1$, with $n$ and $C$ unchanged. Either the benefits per average step are higher, or energy per average step is lower, compared to an ideal network. If the relative benefits per step increases, the relative benefit of the network increases. For $n$ and $C$ unchanged, the only way the network benefit can increase is if $L$ is smaller than $e$. An analogous argument implies that when (energy/benefits) $> 1$, $L$ is greater than $e$. For example, in social networking, the `six' in six degrees of separation may reflect the greater amount of energy required to connect to remotely located people, and the smaller social benefits received from remotely located people, compared to those closer.

Though energy scaling leads to a flattened hierarchy for an ideal network, it may be possible that a physically observable energy hierarchy indirectly manifests itself in real networks of cells in organisms, buildings in a city, or stars in a galaxy.

Equation (\ref{Eq Net20.10}) has a form similar to that for entropy used in information theory, and so may be called, by analogy, the entropy of a network. In 1948, C. E. Shannon derived an equation for the entropy of a set of probabilities\endnote{ C. E. Shannon and W. Weaver, The Mathematical Theory of Communication. (University of Illinois, Chicago, 1949).},
\begin{equation}\label{Eq Net20.40}
	H_r(S) = K \sum^{n}_{i=1} p_i \log_r p_i,
\end{equation}

\noindent to analyze strings of symbols. He called $H$ (the Greek letter eta) in Equation (\ref{Eq Net20.40}) entropy because it has the same form as that used for entropy in statistical mechanics. The $r$ is an arbitrary base of the log, $S$ is the symbol source, $K$ is an arbitrary positive constant, and $p_i$ is the probability of the $i^{th}$ symbol. In Shannon's derivation, probability and information are related. If the probability of an event occurring or not occurring is 100\%, no new information is acquired after its occurrence. Only resolution of uncertainty adds information. In Equation (\ref{Eq Net20.40}), the base of the log is usually 2 because Equation (\ref{Eq Net20.40}) is mostly used in connection with digital communication. $K$ is usually set to 1. 

Like Equation (\ref{Eq Net20.40}), the formula for an ideal network's entropy can be derived using probability. Equality of nodal capacities implies that the average probability that a node in N is an information source is $1/n$. When $p_i = 1/n$, Equation (\ref{Eq Net20.40}) reduces to $K \log_r(n)$, with the base of the log $L$ and the constant $K$ the clustering coefficient, for the reasons stated above. Weighted probabilities and energy scaling both lead to the same formula for network entropy. Each derivation likely implies the other: weighted probability paths imply scaling when $p_i = 1/n$, and scaling implies weighted probability paths. Each describes a different aspect of entropy. An ideal network has maximal uncertainty (or equality) $p_i = 1/n$ for all nodal sources. The resulting equality of nodal capacities leads to energy scaling, maximally efficient and maximally uncertain or equal. In information theory, the joint entropy of a joint event is less than or equal to the sum of the component entropies.

In information theory, entropy is maximal\endnote{ C. E. Shannon, p. 51; A. Ya Khinchin, Mathematical Foundations of Information Theory (Dover, New York, 1957), p. 41.} for a network of n nodes when $p_i = 1/n$. Equivalently, network entropy is maximal if we suppose the energy requirements of $N$'s nodes are equal, or if we scale $N$'s energy by $L$. Why $L$ scales $N$'s energy gives some insight into the operation of a network. Suppose a given signal can be propagated from a proper subset of $N$ consisting of $n/(L^\eta)$ nodes. This is efficient for $N$, because $N$ does not have to use all its nodes' energy\endnote{ On a network's efficiency: V. Latora,, and M. Marchiori, Phys. Rev. Lett. 87, 198701-2 (2001); R. Ferrer i Cancho, and R. V. Sol\'e, PNAS 100: 788 (2003).} any time a signal is to be sent to all or part of $N$. If the speed of the signal is less than $L$ steps per $L$ time units the signal can not reach the whole of the network within $L$ time units; the signaling nodes in the subset are using less than the average amount of energy per node, and the entropy of $N$ is therefore less than optimal. On the other hand, if the speed of the signal is greater than $L$ steps per $L$ time units, the signaling nodes in the subset are using more than the average amount of energy per node, and the entropy of $N$ will also be less than optimal because $N$'s other nodes will have less than the average capacity to transmit. To optimize network entropy a conservative approach is to structure $N$ so that $N$'s nodes have equal capacity to access $N$'s energy, because potentially each node has an equal capacity to benefit $N$. The distribution of equal capacity may occur in some networks naturally due to the randomness of energy distribution.

While nodal self-interest would result in a node tending to accumulate as much energy to itself as possible, networking leads to the emergence of a network benefit, which benefits nodes individually and collectively, and therefore restrains the accumulation of energy by individual nodes. $L$ equaling $e$ reconciles self-interest and the benefit of networking. Since a network is self-similar, the conflict between nodal self-interest (leading to unrestrained accumulation of energy) and network benefit (leading to equal distribution of energy) would arise in cluster generations as well. 

An ideal network maximizes efficiency as a consequence of its assumed attributes. A real network maximizes its energy efficiency by its continual adaptation to its environment. Since both the ideal and real networks are maximally efficient, the ideal by assumption, and the actual by adaptation, an ideal network may be a reasonable model of a real network with similar attributes. 

If the assumptions of an ideal network apply to economic actors, a communication system, bodies that are mutually gravitationally attractive, or a group of molecules, the network will be maximally efficient when the capacities and energy of the network are equally distributed among its nodes. This inference omits consideration of the impact that the network may have on its environment (externalities), and the effect of changes in the environment on $N$. 

Shannon also observed\endnote{ C. E. Shannon, p. 53.} that, for symbols, 

\begin{equation}\label{Eq Net20.80}
	H' = mH.
\end{equation}

 This applies, analogously, to networks. If $H'$ is the rate of a network process, and $H$ is the network's entropy for that process, then $m$ is the process rate when $\eta = 1$, that is, when hierarchical structure and networking began. If the process grows exponentially (which scaling suggests can occur), we can calculate the average rate at which the number of nodes grows, if their number at the beginning of the process (time $t_1$) and at its end (time $t_2$) are known, by solving for $m$ in $n(t_2) = n(t_1)e^{mt}$, where $t = t_2 - t_1$. If the entropy $H$ of a system $S$ at $t_2$ and of its ancestral system at $t_1$ are both known, and $t = t_2 - t_1$, solving $m$ in
\begin{equation}\label{Eq Net30.20}
	H(S(t_2)) = H(S(t_1))e^{mt} 
\end{equation}

\noindent may give an estimated average rate of growth for the entropy itself. The productivity rate of society when $\eta$ was 1 measures society's capacity to use energy before that capacity was multiplied by clustering in a scaled way (i.e. by entropy). With that, knowing the rate of change permits one to date a beginning of a process, because the ending and starting rates of energy utilization, and the degree of energy clustering, are all indirectly known. 

We can use the average rate of growth in the number of nodes or in the size of entropy to estimate when a network's entropy growth began: that is, when $\eta$ was 1. Suppose entropy and the average rate of growth in the number of nodes at a process's beginning $t_0$ and their number at the process's end $t_2$ are all known. Then we estimate the duration of the process by solving for $t$ in $e^{m\eta t} = n(t_2)$, with $t = t_2 - t_0$. Similarly, the average rate of growth in entropy can be used to estimate when $\eta$ was 1. For example, the finding of the age of mitochondrial Eve using DNA may be finding the age of the cluster generation for $\eta  =1$ for diverging mitochondrial DNA; thus Eve would be a representative individual from that cluster generation, not necessarily a single person as appears to be sometimes inferred.

Entropy dating is accurate only if the calculated average rate prevailed for the entire period preceding the earlier of the two dates used for calculation. For example, if neuronal physiology since language began has not changed, then neuronal energy use per step has not changed, and m for lexical growth may have been unchanged during language's development. On the other hand, over millions of years neuronal physiology and the rate of energy supplied by the environment may have varied, and using m for a long period preceding the time for which its average value was determined may yield uncertain results. 

The following observation about conceptual networks applies to the lexical growth example below. Each person in a society possesses networks of ideas; living individuals network with inherited ideas. Suppose that, on average, each person possesses the capacity to access the same concepts. To calculate the entropy of concepts promulgated by the society for a given era, multiply the entropy of that society times the entropy of the concepts that are held in common. The network of ideas common to each average member of a society is like an infrastructure (in a mathematical derivation, a constant). Infrastructures include realized ideas such as roads, buildings, and technologies.

To apply Equation (\ref{Eq Net20.10}) to a real network, the real network's attributes must be similar to those of an ideal network. Then only $n$, $L$ and $C$, which provide statistics about the macrostate of the real network, are needed. Even though nodes permute among clusters for some real networks, the averaging used to calculate $L$ and $C$ for a real network in effect assigns to clusters distinct nodes of equal average capacities.

Researchers' calculations have enabled them to estimate the path length for real networks, such as, for example, a human brain (2.49)\endnote{ S. R. Achard, R. Salvador, B. Whitcher, J. Suckling, and E. Bullmore, The J. of Neuroscience 26(1), \textbf{63} (2006). They found $C = .53$.}, the nervous system of the worm C. elegans (2.65)\endnote{ D. J. Watts and S. H. Strogatz.}, and the English lexicon (2.67)\endnote{ R. Ferrer i Cancho  and R. V. Sol\'e, Pro. R. Soc. B, \textbf{268}, 2261 (2001). $L = 2.67$,$ C = .437$ based on 3/4 of the million different words of the British National Corpus (about 70 million words). A study of the English lexicon based on words in an online thesaurus, likely less representative of English usage is: A. Motter and A. de Moura, Y. Lai, and P. Dasgupta, Phys. Rev. E. \textbf{65} 065102(R) (2002). They obtain $L = 3.16$, $C = .53$, which would give $\eta = 6.14$.}. For these examples, $L$ is close to $e$, 2.71828. Perhaps in these examples the conflict between nodal self-interest and the benefit of networking has been efficiently reconciled.

We now estimate the effect of adding nodes to a network. Let $H_1'$ be the rate for a network process for a network of $n_1$ nodes. Let $H_2'$ be the rate for a larger number of nodes $n_2 = (n_1 + A)$. Assume $L$, $C$ and $m$ do not change as the network grows. Then the increase in $H_1'$ due to $A$ additional nodes is 

\begin{equation}\label{Eq Net30.40}
\begin{split}
	H_2' - H_1' &= mC \log_L(n_2) - mC \log_L(n_1) \\
	&= mC \log_L(n_2/n_1)\\
	&= mC \log_L(1 + A/n_1). 
\end{split}
\end{equation}

If $A = 1$, Equation (\ref{Eq Net30.40}) represents the difference that the presence or absence of an individual makes to a group. If $n_1$ is small, likely $C$ is closer to 1 and $L$ smaller than for a large group, and an individual makes a larger difference to the entropy of the group. A related issue arises in the early 1980s proposed estimate, dubbed Metcalfe's law, that the profitability of a commercial communication network grows with the square of its size. Equation (\ref{Eq Net30.40}) may apply instead\endnote{ A. Odlyzko and B. Tilly of the University of Minnesota, http://www.dtc.umn.edu/~odlyzko /doc/metcalfe.pdf (2005) in A refutation of Metcalfe's Law and a better estimate for the value of networks and network interconnections; B. Briscoe, A. Odlyzko, and B. Tilly, IEEE Spectrum, July 2006, 26. They estimate that the value of a communication network of size $n$ grows like $n\log(n)$.}. Since the entropy of a large network changes slowly with $n$, much of the commercial benefit of adding customers to a large network likely results from economies of scale. For merging related existing networks, the joint entropy is less than the sum of the component entropies if the processes of the two are not independent, as may be the case, for example, for fixed line and cellular telephone networks. 

As an example of entropy dating, suppose that humans' lineal ancestor had one third as many neurons 3 million years ago. Then $H(early \ brain)$ would be 14.077, compared to $H(modern \ brain) = 14.71$\endnote{ Using $L$ and $C$ from S. R. Achard, R. Salvador, B. Whitcher, J. Suckling, and E. Bullmore for $\eta(modern)$, and for $\eta(earlier \ brain)$, and assuming the earlier brain had one third the neurons, where $n$ is the number of neurons. Assuming $n = 10^{11}$ neurons, from J. G. Nicholls, A. R. Martin, B. G. Wallace, and P. A. Fuchs,  From Neuron to Brain (4th ed.) (Sinauer, Sunderland, Mass., 2001), p. 480.}. The average growth rate in neuronal entropy over 3 million years would be .01478\ldots per million years. At that rate, it would take 995 million years for neuronal entropy to evolve from 1 to 14.71, or from the first connected neurons to $10^{11}$ neurons. This manner of estimation requires that the energy requirements, the energy supply, and the capacity of neurons were on average the same over the whole period of their development, probably unlikely given the number of years involved, though if networked neurons optimized their $L$ and $C$ early in their development, the values of $L$ and $C$ may have changed only slightly over those years. 

The estimated 1989 entropy of 350 million English speakers (a social network)\endnote{ Using $L = 3.65$, $C = .79$ for 225,226 actors from Watts and Strogatz.} is 12 and of an English lexicon (a conceptual network)\endnote{ Using $L$ and $C$ for English from Ferrer i Cancho  and Sol\'e (2001).} of 616,000 words\endnote{ Oxford English Dictionary (OED) at 1989.}, 5.93. The entropy of English lexical growth is the product of the two entropies. We now wish to estimate the average basal rate lexical growth, the rate of lexical growth without the multiplier effect. 

The estimated 1657 entropy\endnote{ Again using $L = 3.65$, $C = .79$ for 225,226 actors from Watts and Strogatz.} of 5,281,347 English speakers\endnote{ E.A. Wrigley, R. Schofield \& R.D. Lee. The population history of England, 1541-1871: a reconstruction Cambridge University Press, 1989, Table 7.8, following p. 207, for the year 1656.} is 9.445. The entropy of the 1657 English lexicon of 200,000 words is 5.431. The product of the average population entropy of 1657 and 1989 times the average lexicon entropy for 1657 and 1989 is $10.72 \times 5.68 = 60.94$. This is the average value of the multiplier for the period from 1657 to 1989. Using this multiplier, the basal lexical growth rate from 1657 to 1989 is about 5.6\% per thousand years.

A independent means of checking the 5.6\% per thousand year rate involves glottochronology\endnote{ What is Glottochronology, p 271, in M. Swadesh The Origin and Diversification of Language. (Aldine-Atherton, Chicago, 1971).}. Glottochronology uses the rate at which two related languages diverge to date their common ancestral language. In the 1960s, Morris Swadesh determined that after 1,000 years, two related Indo-European languages shared on average 86\% of the words on a Basic List he compiled (i.e. a 14\% divergence after thousand years)\endnote{ M. Swadesh, p. 276.}. The divergence between two related languages after a thousand years, if now adjusted by recent work by Gray and Atkinson\endnote{ R. D. Gray and Q. D. Atkinson, Nature (London) \textbf{426}, 435 (2003), estimate Indo-European at 8,700 years ago. Swadesh, 37 years before Gray and Atkinson, estimated Indo-European beginning at least 7,000 years ago (p. 84). I assume Gray and Atkinson's estimate is an improvement on Swadesh's, and so multiply 14\% by 7037/8700 to obtain 11.32\%.}, is about 11.32\% per thousand years\endnote{ R. D. Gray and Q. D. Atkinson, Nature (London) \textbf{426}, 435 (2003), estimate Indo-European at 8,700 years ago. Swadesh, 37 years before Gray and Atkinson, estimated Indo-European beginning at least 7,000 years ago (p. 84). I assume Gray and Atkinson's estimate is an improvement on Swadesh's, and so multiply 14\% by 7037/8700 to obtain 11.32\%.}. If each of the two daughter languages diverges from the mother language at the same average rate, then the average rate of divergence per daughter language is one half of 11.32\% per thousand years, which is 5.66\% per thousand years, very close to the 5.6\% per thousand years found using the entropies of the English speaking population and English lexicon. 

If we assume that the English lexical growth rate is representative of lexical growth rates and that human lexical growth is a stable capacity, we can use the 5.66\% per thousand years basal lexical growth rate to estimate when language began. We assume that ancestral societies, consisting of 50 individuals\endnote{ R. Dunbar, Grooming, Gossip and Language. (Harvard University Press, Cambridge, Massachusetts, 1997),  p. 120 - 123.} using 100 different call signals immediately preceded language's beginning, and had modern values for $L$ and $C$ for their society. It would take about 154,000 years for the lexicon to grow from 100 words to the 616,500 words of the OED in 1989 at the rate of 5.66\% per thousand years.

In addition to the three problems confronted in growing a language is a fourth problem: choosing, from the menu of concepts and opportunities that a society has stored up in all cluster generations of its language, culture, and economy, which ones best apply to the immediate circumstances. What we regard as individual intelligence may consist to a large extent of learning the conceptual menu created by societies over thousands of years, as seems to be suggested by the multiplicative effect of network entropy.

Some concepts and theorems in information theory may be adaptable to the entropy of a network. Being able to calculate entropy may assist in the analysis of economic\endnote{ If one dollar is a claim on (or a proxy for) one unit of energy, then to maximize the entropy of the economy, the members of society should maximize the efficiency of each dollar used to acquire benefits from society. This requires the economy to permit network adaptation (and therefore, nodal and cluster adaptability) that maintains, for equilibrium states, the equal ratio of one dollar to a unit of benefit.}, biological, communication, conceptual, and social networks. If the entropy of a network has these uses, then statistical information about real networks of interest will be helpful. 

\theendnotes

\end{document}